# ADAPTIVE POSITION-BASED FLUIDS: IMPROVING PERFORMANCE OF FLUID SIMULATIONS FOR REAL-TIME APPLICATIONS

Marcel Köster and Antonio Krüger

German Research Center for Artificial Intelligence, Saarland Informatics Campus, Saarbruecken, Germany

***ABSTRACT***

*The Position Based Fluids (PBF) method is a state-of-the-art approach for fluid simulations in the context of real-time applications like games. It uses an iterative solver concept that tries to maintain a constant fluid density (incompressibility) to realize incompressible fluids like water. However, larger fluid volumes that consist of several hundred thousand particles (e.g. for the simulation of oceans) require many iterations and a lot of simulation power. We present a lightweight and easy-to-integrate extension to PBF that adaptively adjusts the number of solver iterations on a fine-grained basis. Using a novel adaptive-simulation approach, we are able to achieve significant improvements in performance on our evaluation scenarios while maintaining high-quality results in terms of visualization quality, which makes it a perfect choice for game developers. Furthermore, our method does not weaken the advantages of prior work and seamlessly integrates into other position-based methods for physically-based simulations.*

## 1. INTRODUCTION

Modern games leverage real-time physics simulations to provide a realistic experience. In particular, fluid simulations that can interact with other physical objects in the scene have become more and more popular [1,2,3].The approach of Position Based Fluids (PBF) by Macklin et al. [4] has relaxed previous limitations while maintaining stability. It uses position-based dynamics and constraint functions based on Smoothed Particle Hydrodynamics (SPH) [5, 6].Moreover, it allows for a seamless integration into other position-based methods that are widely spread in the scope of real-time applications [7].

However, fluids require a notion of incompressibility (constant fluid density) for a realistic simulation. In the scope of PBF, this requirement is realized by an iterative approach that tries to adjust particle positions to reach the desired incompressibility. This iterative process suffers from low convergence rates when simulating large volumes with a large number of particles. In these cases, the required solver iterations have to be increased significantly, implying a huge impact on the runtime of the overall simulation.

In this paper, we demonstrate a novel adaptive approach for the simulation of fluids using position-based constraint functions, called Adaptive Position-Based Fluids (APBF).Our concept is very easy to integrate into other position-based simulation frameworks and does not impose further restrictions compared to PBF.Furthermore, it allows for considerable increases in performance, which is desirable for games.

Depending on the situation and camera setup during runtime, our method can adaptively adjust the particle positions of the fluid via fine-grained level-of-detail (LOD) information per particle. The LOD information is then used to adaptively adjust the number of solver iterations for each particle in the fluid (Figure 1).This allows to trade simulation precision for performance in different areas. Our approach is also able to achieve high-quality results with respect to rendered fluid volumes. The evaluation itself is based on different scenarios that are mainly compared on the basis of visual quality (visible differences) and performance.

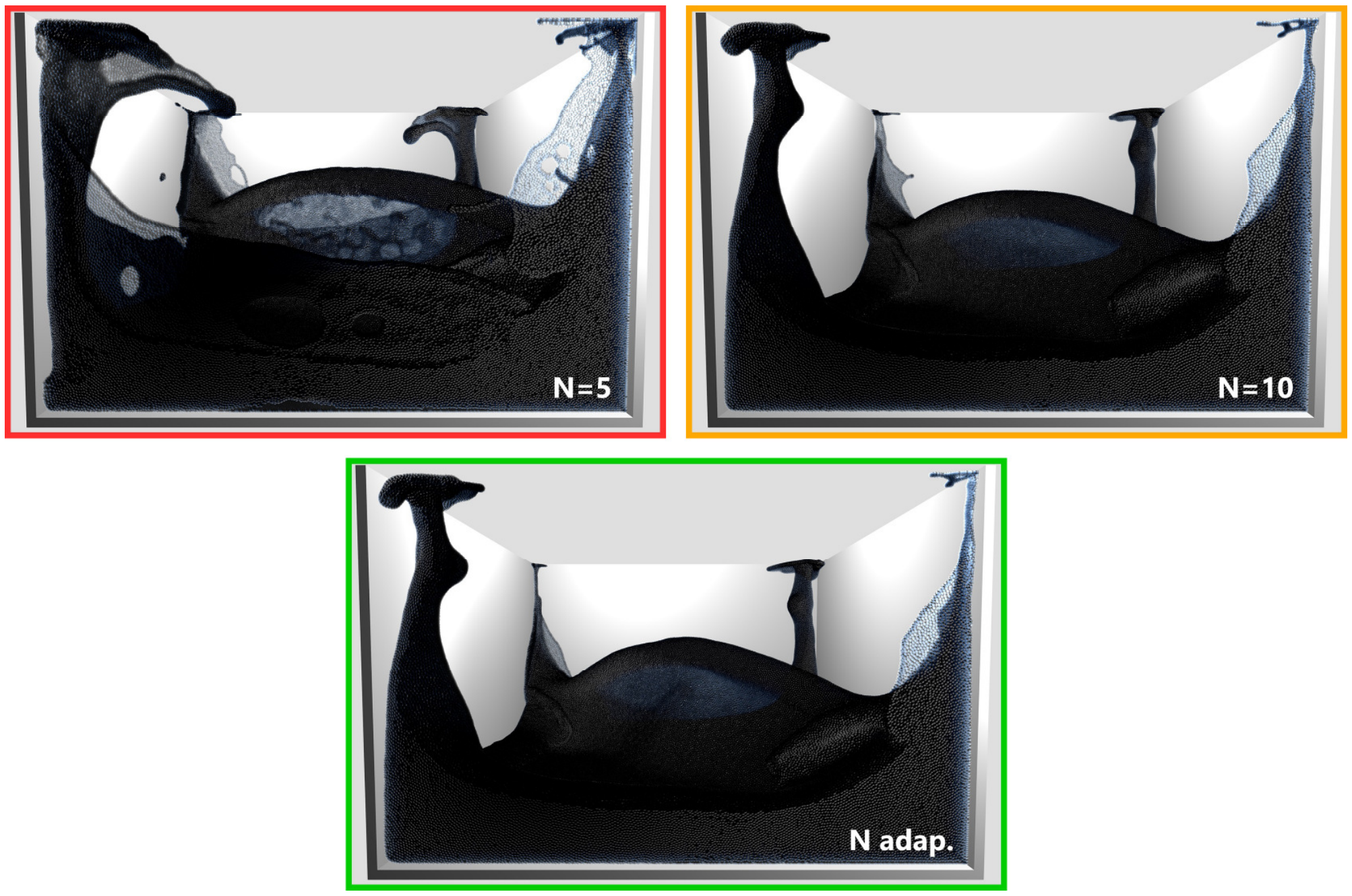

Figure 1. Images from the second evaluation scenario with different methods and number of iterations $N$.The variants in the top row were simulated using Position Based Fluids with $N = 5$ (left, red) and $N = 10$ (right, orange). Our adaptive method is presented in the bottom row (green) and uses $N \in \{5, \ldots, 10\}$.

## 2. Related Work

### 2.1. Fluid Simulations

Müller et al. use SPH-based quantities, which form the basis for the computation of an acceleration vector (force-based) [8, 9].This vector is used to update the particle movement of all particles in every time step. Their approach is responsible for a fairly high amount of compressibility in many situations [10].Premoze et al. [11] solve the Poisson equation iteratively in every time step for particle-based fluids to ensure incompressibility. However, this is problematic for interactive applications as described by Becker et al. [10].The expensive solver approach from [11] can be circumvented by WCSPH [4], which allows for small density variations. However, WCSPH suffers from time-step ($\Delta t$) restrictions.

An alternative to WCSPH is PCISPH [12].PCISPH weakens the time-step restrictions and also achieves good results compared to WCSPH.The authors use an iterative algorithm to predict the next position and next velocity in each iteration.Ihmsen et al. further improved SPH-based fluid simulation in terms of time-step restrictions and performance with IISPH [13].A recent advance in the direction of interactive and incompressible fluid simulations is the paper *Position Based Fluids* (PBF) by Macklin et al. [4].It describes the modelling of incompressible fluids with the help of position-based constraints in the context of the Position Based Dynamics framework by [14, 15].Those constraints are solved iteratively in every time step.

## 2.2. Adaptive Fluid Simulations

Adams et al. introduced an adaptive sampling model, which is based on heterogeneous particle sizes [16].The sampling condition is based on visual importance of affected regions with respect to near object geometry. This allows for a reduction in the number of particles inside a fluid volume. One of the main contributions is the adaptive sampling approach that describes merging and splitting of particles. The decision on merging or splitting is based on the so-called *extended local feature size*. From a high-level point of view, this can be seen as the distance to the surface and next obstacle in the scene.

Hong et al. use a hybrid grid and particle-based method and introduce a multi-layer approach in which the simulation domain is split into four layers (based on the distance to the surface) [17, 18].Related to the previously presented approach, the authors also use different particle sizes in different layers. Every layer has its unique rules on whether to split or merge particles. This approach is also similar to the one by Zhang et al. who also perform splitting and merging based on several conditions [19].

Another adaptive simulation is the *Two-Scale Particle-Simulation* by Solenthaler et al. [20].They are using two differently scaled simulations in parallel, which remove the need for splitting and merging. The larger-scaled simulation acts as a base simulation of the fluid domain. This simplifies neighbour search and related computations. Horvath et al. extended this approach to an arbitrary number of levels with smooth collision-detection support [21].The LOD can be adapted according to camera-dependent properties, such as the distance to the used camera.

Goswami et al. [22] differentiated between active and inactive particles that are subject to different position-adjustment steps. Inactive particles remain in a sleeping state (not moving) until they become reactivated. The decision on the activity state of a particle is based on the observation that not all particles contribute significantly to the overall visual appearance: particles (or particles in their surroundings) that are moving fast or are close to the boundary of the fluid are most important for visual quality.

There have also been a huge variety of approaches to select the time-steps adaptively.Goswami et al. [23] introduced an approach for WCSPH that groups particles into regions. Those receive different time-steps and are updated at different frequencies in order to improve performance.Ihmsen et al. [24] introduced adaptive time-steps for PCISPH, also to improve performance. In contrast to these methods, our approach does not run multiple differently-scaled particle simulations in parallel and does not use different particle sizes or adaptive time steps. Instead, we manipulate the number of solver iterations per particle to improve performance while preserving a high visual quality. The approach most similar to our method is the one by Goswami et al. [22], since it also differentiates between active and inactive particles and uses a similar LOD criterion.

## 3. INCOMPRESSIBILITY IN PBF

Position based fluids rely on the Position Based Dynamics (PBD) principle [15].In general, PBD uses a set of non-linear constraint functions that work on positions $\boldsymbol{p_i}$, their mass $m_i$ and their weighting $w_i = \frac{1}{m_i}$, where $i$ refers to the $i$-th point.These constraint functions are then solved in an iterative manner.In contrast to the PBD concept that uses an inherently sequential Gauss-Seidel solver, PBF leverages a parallel Jacobi-style solver.

PBD tries to find a position-correction configuration $\Delta\boldsymbol{p_i}$ that manipulates the positions in order to satisfy constraints of the form [14]:

$$C_i(\boldsymbol{p} + \Delta\boldsymbol{p}) = 0 \text{ and } C_i(\boldsymbol{p} + \Delta\boldsymbol{p}) \leq 0, \tag{1}$$

where $C_i$ refers to the $i$-th constraint and $\boldsymbol{p}$ describes the concatenated vector of positions.In order to solve for the position-correction configuration $\Delta\boldsymbol{p}$, we can approximate $C_i$ via Taylor expansion:

$$C_i(\boldsymbol{p} + \Delta\boldsymbol{p}) \approx C_i(\boldsymbol{p}) + \nabla_{\boldsymbol{p}} C_i(\boldsymbol{p}) \cdot \Delta\boldsymbol{p}. \tag{2}$$

By choosing the correction to be in the direction of the gradient $\nabla C_i$ and weighting according to the masses, we receive

$$\Delta\boldsymbol{p} = \lambda w_i \nabla_{\boldsymbol{p}} C_i(\boldsymbol{p}), \tag{3}$$

with $\lambda$ being a scaling factor along the gradient.Following the rewriting steps in [15], we receive the formula for the scaling factor

$$\lambda_i = -\frac{C_i(\boldsymbol{p})}{\sum_j w_j \left|\nabla_{\boldsymbol{p}_j} C_i(\boldsymbol{p})\right|^2}. \tag{4}$$

PBF uses this general concept with the help of a *density constraint* by [25]:

$$C_i(\boldsymbol{p}) = \frac{\rho_i}{\rho_0} - 1 = 0, \tag{5}$$

with $\rho_0$being the rest density of the fluid and $\rho_i$ the current mass density.From a PBD's point of view, this equality constraint is inserted for every particle in the simulation, which results in an evaluation of $\rho_i$ at the location of every $i$-th particle.For this purpose, we can leverage the default SPH-based density estimator [26]:

$$\rho_i = \sum_j m_j W(\boldsymbol{p}_i - \boldsymbol{p}_j, h), \tag{6}$$

where $W$ is the smoothing kernel and $h$ is the smoothing length of the kernel.In our scenes, we typically choose a fixed smoothing length $h$, and kernels $W$ with finite support.The gradient of the density constraint based on SPH can be computed using the general SPH-based gradient formulation by [27].Using $\lambda_i$, we can formulate the position correction $\Delta\boldsymbol{p_i}$ as

$$\Delta\boldsymbol{p_i} = w_i \frac{1}{\rho_0} \sum_j (\lambda_i + \lambda_j) \nabla W(\boldsymbol{p}_i - \boldsymbol{p}_j, h). \tag{7}$$

By applying the position correction $\Delta\boldsymbol{p_i}$ to all particles in every iteration, we can propagate the influences of the density constraint to all neighboring particles.This implies a low convergence rate with a large number of particles, since the corrections are only propagated to the direct neighbors in every iteration.

## 4. ITERATION-ADAPTIVE POSITION-BASED FLUIDS

A higher number of solver iterations implies a better approximation of the used density constraint, and thus, of the desired incompressibility. The number of iterations, however, significantly influences the runtime of the simulation. The general idea is straight forward: Choose the number of iterations adaptively. Doing so results in a lower or higher approximation of incompressibility depending on the situation. For this reason, a fine-grained approach is required, allowing an adjustment on a per-particle basis. First, we require a notion of a particle level that represents the level of detail per particle (LOD).This level information is defined as

$$lv(p_i) : \mathbb{N} \rightarrow \{1, \dots, N_l\} \subset \mathbb{N}, \tag{8}$$

where $p_i$ refers to the position of the $i$-th particle and $N_l$ to the maximum LOD.

This definition is somewhat similar to other adaptive approaches. All particles with a certain level that is greater or equal to the currently processed one have to be considered at the same time during a solver step. Otherwise, a particle with a higher LOD would only be considered in a specific solver step in the future, and misses the previous adjustment steps for the rest of the particles. This would lead to instability and too-coarse-grained approximations (Section 4.3.). Hence, we will call a particle with index $i$ with respect to a given solver iteration $l$ *active* if it has to be considered in the current iteration of the solver loop:

$$ac(p_i, l) := lv(p_i) \geq l, \tag{9}$$

where $l$ is $\in \mathbb{N}$.For simplicity, we assume that the level can be directly mapped to the number of iterations.In order to reason about all particles that are active with respect to a specific iteration $l$, we define the set $P_l$ as

$$P_l := \{p_i | ac(p_i, l)\}. \tag{10}$$

Particles that are already finished and do not need to be considered in future iterations are given by

$$\widehat{P_l} := \begin{cases} \emptyset, & l \leq 1 \\ P_1 \setminus P_l, & l > 1. \end{cases} \tag{11}$$

A visualization of these definitions can be found in Figure 2.

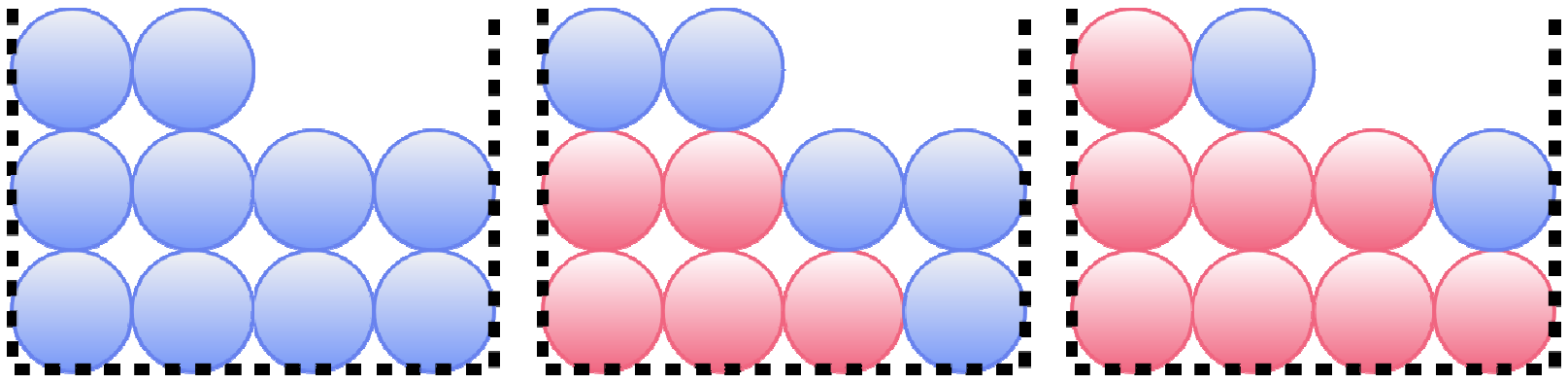

Figure 2. Schematic visualizations of three adaptive iterations (from left to right: iterations 1 to 3). Particles in red will not be considered for further iterations and remain fixed at their position. Blue particles will still be considered and adjusted accordingly. For instance, the right particle in the top row has at least a LOD of 3, since it is considered in all shown iterations (1 to 3).

## 4.1. Adaption Models

The required LOD information per particle can be computed using an analysis based on particle data or rendering information. We choose the rendering-feedback approach to reduce additional overhead by reusing information that is used for visualization purposes. We have integrated and evaluated two possible approaches to compute the LOD information per particle. The first approach uses the distance to the camera (DTC) of every particle, which can be computed using the particle locations and the camera transformations (Figure 3).

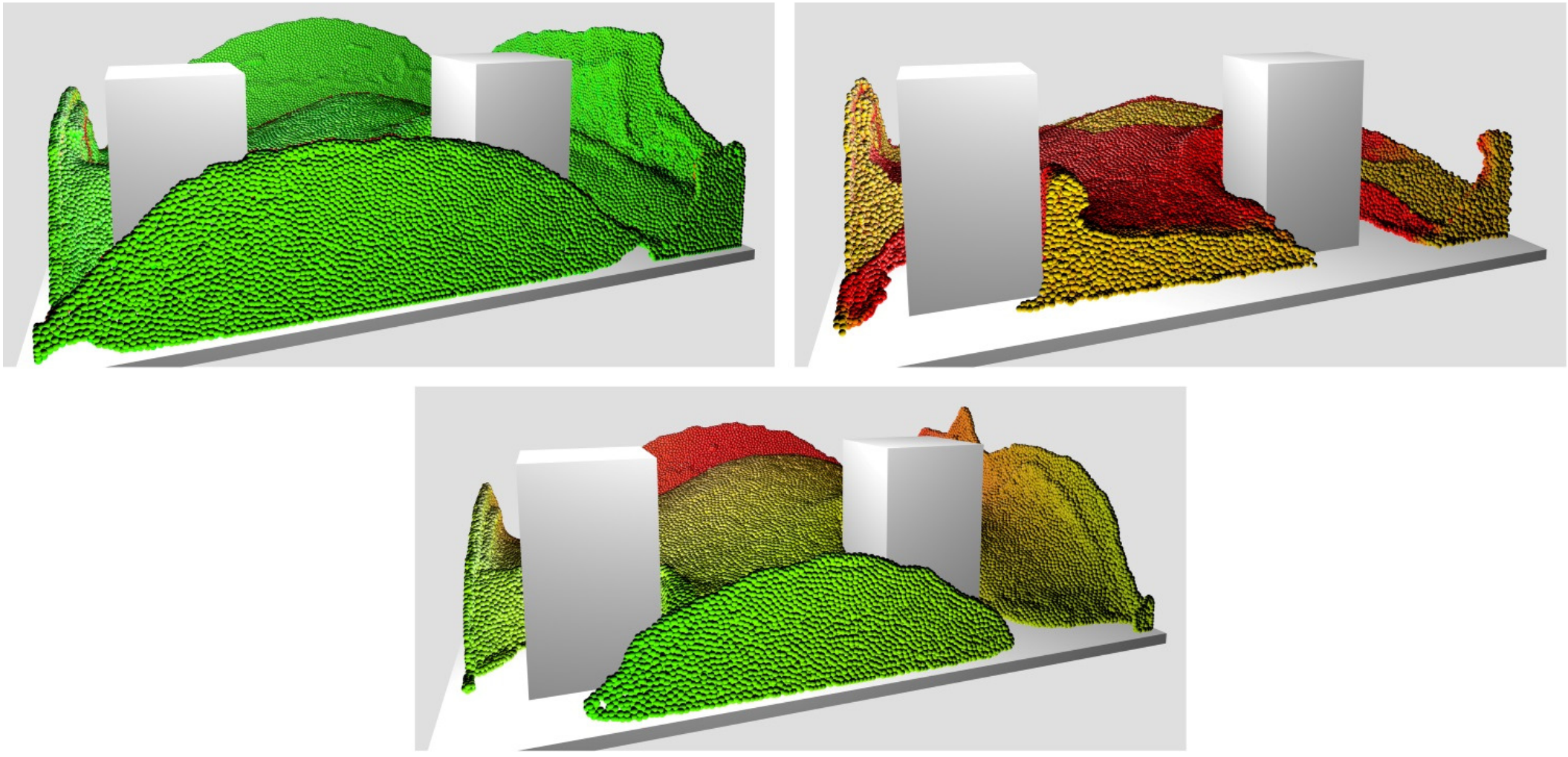

Figure 3. Sample LOD visualizations of the first evaluation scenario for different APBF models. Top left: high-LOD particles classified by DTVS. Top right: low-LOD particles classified by DTVS. Bottom: DTC classification.Color coding**:** green, high LOD; yellow, lower LOD; red, low LOD.

The second model realizes a camera-based distance-to-the-visible-surface approach (DTVS).In comparison to the actual distance of a particle to the surface of the fluid volume, we will treat particles as surface particles if they are directly visible from the camera (Figure 3).We then compute the distance of a particle to its nearest surface particle along a ray from the camera. Figure 4 shows a schematic classification according to the DTVS approach. Dynamic changes of the camera position are implicitly reflected in the LOD computation, since the LOD per particle is updated in every frame.

The actual LOD information is determined from the distances in both cases via linear interpolation between the maximum and minimum distance and the lowest and highest LOD.For customization purposes of the simulation quality, the distances for the actual interpolation can either be user-defined or automatically resolved. Moreover, multiple camera perspectives in parallel can be handled by blending multiple LOD information per particle.

## 4.2. Incompressibility

When adjusting the number of iterations adaptively, the worst-case approximation is the one with the lowest LOD, which still involves a single simulation step in general. In the best case, the approximation is equal to maximum LOD.The computed LOD-distribution depends on the used LOD model (Section 4.1.) and how it influences a particular region of the simulation.

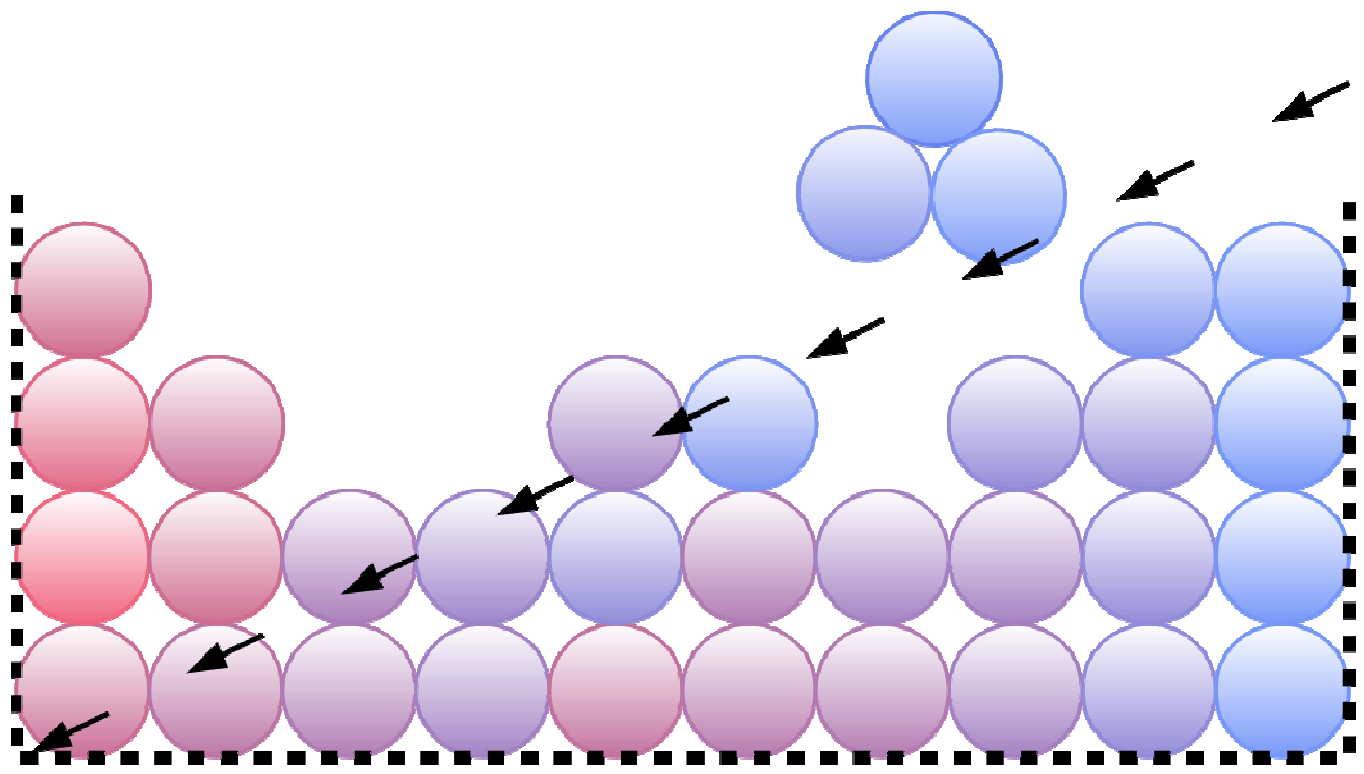

Figure 4. Sample classification according to the DTVS model and the given viewing direction (indicated by black arrows). High-LOD particles are bluish and low-LOD particles are reddish. Note that linear interpolation of the distances avoids significant differences of LOD assignments to neighboring particles.

Related to the density distribution, there can be regions with a lower density and regions with a higher density. However, the average density across the whole simulation domain remains comparable to the computed average density of PBF.This is caused by our solver approach, where particles with a higher LOD can compensate for over- and under-approximations of the density from particles with lower LOD in further iterations (Figure 5).Using the presented adaption models, the integrated linear-interpolation scheme between the different LOD allows for an even distribution of the LOD across the whole simulation domain (Figure 4).Therefore, a sufficient amount of particles can always compensate for coarse approximations from lower-LOD particles in our scenarios. Changing the LOD-computation strategy to a non-evenly distributed LOD assignment (e.g. non-linear LOD assignment from the raw depth buffer) can cause noticeable changes in the average density and the visualization.

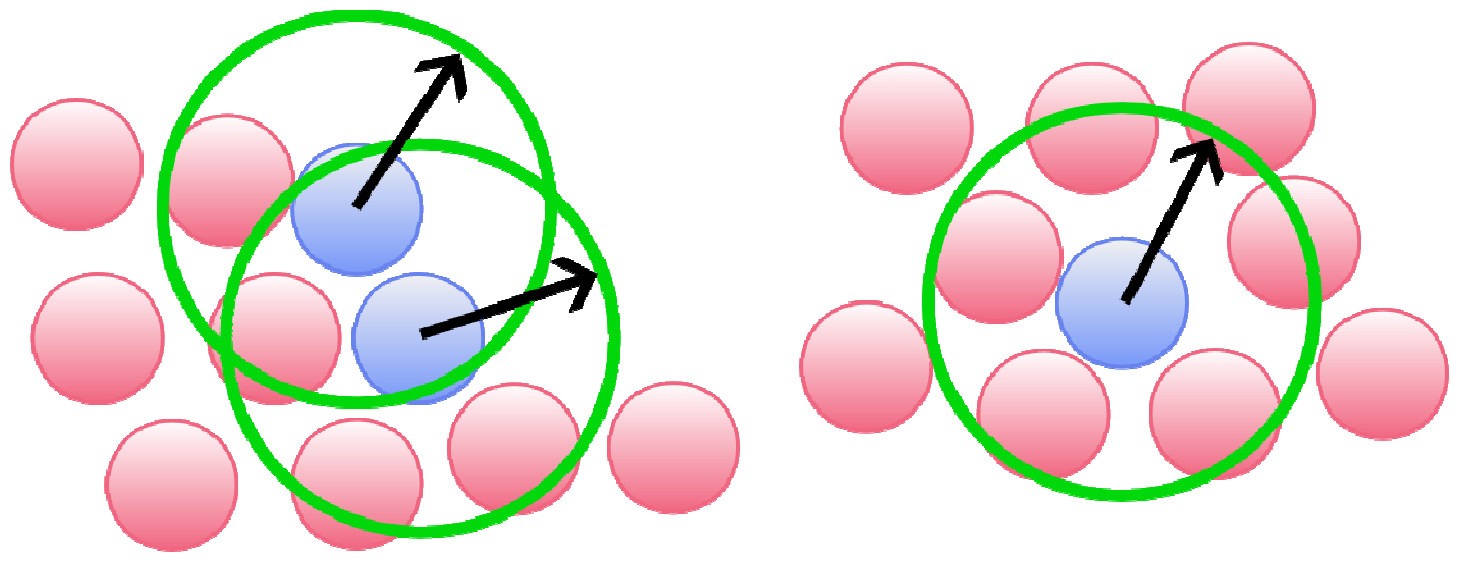

Figure 5. Adjustment of active particles (blue) in the neighborhood of inactive particles (red) and active ones. The green circles refer to the SPH-based smoothing radii and the black arrows indicate the direction vector of $\Delta\boldsymbol{p}$. In this way, active particles can compensate coarse-grained approximations of the density of inactive ones that lie in their smoothing radii.

### 4.3. Stability and Robustness

PBF is stable after a single simulation step assuming a valid configuration in terms of time-step size number of iterations. Every further simulation step does not introduce or cause instability during a simulation run.APBF explicitly manipulates possible further iterations and adjustments per particle, which implies modified $\Delta\boldsymbol{p}$ updates in each other iteration.Inactive particles from the set $\widehat{P}_l$ will not be updated any more after iteration $l-1$, and their position remains fixed.Position updates of active particles are adjusted according to the other active and inactive ones (Figure 5).Since all particles in the set $P_l$ are always considered at the same time in iteration $l$, particles in $P_l$ cannot cause instability with respect to each other (see PBF method).Since those particles also cannot introduce instability with particles in $\widehat{P_l}$, no particles in $P_l$ can introduce instability. This also implies that particles transitioning from low LOD to high LOD and vice versa cannot introduce instability.

However, low-LOD particles can become poorly adjusted in relation to non-fluid particles. This rough approximation of position adjustments for a subset of particles in the simulation domain can cause insufficiently solved contact constraints (Figure 6).Performing the default solver steps on these non-desired particle locations will cause additional velocity to be added to the affected particles, which in turn causes visual artifacts to appear. In order to compensate for these artifacts, we apply the concept of pre-stabilization [7] to a subset of the particles.Pre-stabilization in our case moves particles out of obstacles by adjusting the predicted as well as the current positions before the actual constraint-solving step (Figure 6).

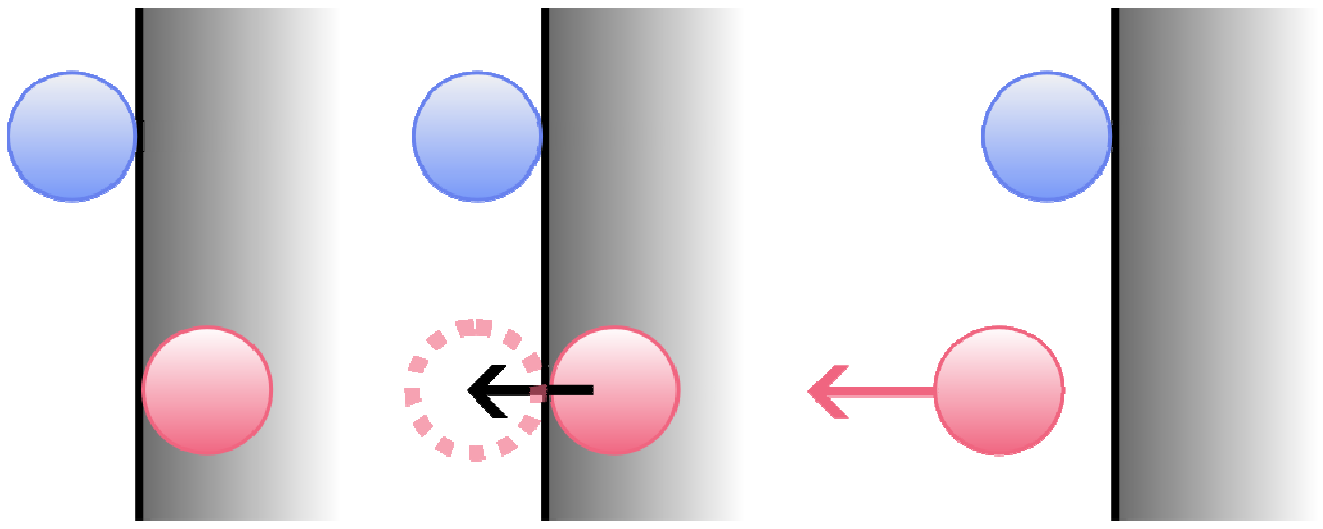

Figure 6. A high-LOD (blue) and a low-LOD particle (red) at the beginning of a frame (left). Due to coarse-grained approximations, the low-LOD particle was not completely moved out of the obstacle. The applied position corrections in the next step will move the red particle to its predicted location (center). Hence, it will be accelerated out of the obstacle, which causes visible artifacts (right).

## 5. ALGORITHM

The main simulation algorithm is shown in Figure7.Regions in red indicate changed parts in comparison to the PBF algorithm, and green regions (line 7 and lines 9 to 11) indicate new parts. We first perform an application of external forces $f_{ext}$ and a prediction of new positions for all particles in $P_l$.Afterwards, we discover neighboring particles and contact collisions (with obstacles) for all particles in $P_1$.We apply a user-defined number of pre-stabilization iterations for all contact constraints that manipulate particles in the set of $\widehat{P_S}$, where $S$ is also a user-defined value between 1 and $N_l$.

Inside the main simulation loop, we perform all operations on active particles in the set $P_{iter}$.Contact constraints that affect the current particle $i$ will be solved during the calculation of $\Delta\boldsymbol{p}_i$.In the final step, all particles in $P_1$ will be modified according the concept of PBD.

## 6. Implementation Details

We realized our adaptive fluid-simulation solver in C++ AMP for GPUs [28].We use signed-distance fields for collision detection and leverage the counting-sort based approach by [29] for efficient neighbor discovery on GPUs, in contrast to other methods for CPUs [30].Since particles are reordered during the neighbor-discovery phase, memory coherence during the evaluation of the density constraint is greatly improved.

The adaption models are based on renderer information that is retrieved by analyzing the depth image of the scene without occlusions by scene geometry. The depth information can be directly resolved by rendering all particles as splatted spheres, which is also used by our and other common screen-space based rendering approaches for particle fluids [2, 31].Furthermore, a single pass over all particles is used to compute and propagate the LOD information.

**for all** particles $i \in P_1$ **do**
  apply forces $\boldsymbol{v_i} \Leftarrow \boldsymbol{v_i} + \Delta t\ f_{ext}$
  predict position $\boldsymbol{x_i^*} \Leftarrow \boldsymbol{x_i} + \Delta t\ v_i$
**end for**
**for all** particles $i \in P_1$ **do**
  find neighboring particles $N_i(\boldsymbol{x_i^*})$
  find contacts for pre-stabilization
**end for**
**while** iter<stabilIterations **do**
  perform contact responses for contact $C_i$ with $C_I \in \{\ C\ |\text{particles}(C) \in \widehat{P_S}, \}$
**end while**
**while** iter< $N_l$ do
  **for all** particles $i \in P_{iter}$ **do**
    calculate $\lambda_i$
  **end for**
  **for all** particles $i \in P_{iter}$ **do**
    calculate $\Delta \boldsymbol{p_i}$
    perform contact responses
  **end for**
  **for all** particles $i \in P_{iter}$ **do**
    update positions $\boldsymbol{x_i^*} \Leftarrow \boldsymbol{x_i^*} + \Delta \boldsymbol{p_i}$
  **end for**
**end while**
**for all** particles $i \in P_1$ **do**
  update velocity $\boldsymbol{v_i} \Leftarrow \frac{1}{\Delta t}(\boldsymbol{x_i^*} - \boldsymbol{x_i})$
  update position $\boldsymbol{x_i} \Leftarrow \boldsymbol{x_i^*}$
**end for**

Figure 7. APBF-simulation loop based on PBF

# 7. EVALUATION

The overall evaluation is based on the pure version of PBF and our adaptive version and focuses on the visual quality of the simulation: Selected frames of the PBF simulation are visually compared to the APBF versions. Rendering of fluids relies on direct semi-transparent particle visualization in the form of non-distorted spheres in order to properly compare certain regions in the images. Furthermore, we do not apply any smoothing operations (e.g. Laplacian smoothing) to the particle data before rendering and make use of a single camera in every scenario.

The evaluation is performed on three scenarios, which make use of differently configured fluid setups. The fluid and scene configuration is then fixed for all performance and visual-quality evaluations for PBF and APBF.All scenarios use a frame time of $\Delta t = 0.0016$ seconds andtwo time-steps per frame. For every scenario, a minimum (required for an acceptable result) and a default number of iterations (for an appealing result) based on PBF are determined. The number of iterations was determined by analyzing the density information to ensure a proper representation of the modeled fluid. An appealing result achieves a reasonable density distribution across the whole simulation domain, in contrast to the much coarser approximation in the case of an acceptable result. Note that an increase in the number of iterations also implies an increase in apparent viscosity in our scenarios, which is an inherit side effect of PBF.This, however, can be avoided by using an unilateral density constraint in combination with force-based cohesion viscosity [7, 32].

Performance tests were executed on two different GPUs from different vendors: a GPU from NVIDIA (GeForce GTX 680) and a GPU from AMD (Radeon HD7850).A performance measurement is the median execution time of 1000 simulation steps and 100 application executions (see Table 1).

Table 1. Performance results for the presented evaluation scenarios on an NVIDIA GeForce GTX 680 and an AMD Radeon HD7850

| **Scenario** | **Particles** | **Configuration** | **Iterations** | **Frame time in milliseconds** | | **Performance Improvement** | |
|---|---|---|---|---|---|---|---|
| | | | | NVIDIA | AMD | NVIDIA | AMD |
| Dam Break | 216k | PBF | 3 | 20 | 25 | - | - |
| | | PBF | 6 | 39 | 46 | - | - |
| | | APBF (DTC) | {3,…6} | 24 | 30 | 63% | 53% |
| | | APBF (DTVS) | {3,…6} | 23 | 29 | 70% | 59% |
| Double Dam Break | 673k | PBF | 5 | 194 | 238 | - | - |
| | | PBF | 10 | 410 | 476 | - | - |
| | | APBF (DTC) | {5,…10} | 241 | 290 | 70% | 64% |
| | | APBF (DTVS) | {5,…10} | 232 | 281 | 77% | 69% |
| Multi Dam Break | 225k | PBF | 4 | 28 | 35 | - | - |
| | | PBF | 8 | 54 | 67 | - | - |
| | | APBF (DTC) | {4,…8} | 38 | 48 | 42% | 40% |
| | | APBF (DTVS) | {4,…8} | 34 | 46 | 59% | 46% |

## 7.1. Dam Break

The scene consists of a single dam-break location of a water-like fluid, which is released after the start. It contains 216,000 particles and uses a number of iterations $N = 3$ for an acceptable density approximation and $N = 6$ for an appealing result in terms of density.

Figure 8 shows the visualized simulation during the collision of the fluid with the walls. In the case of PBF with $N = 3$, many particles are only loosely connected to their neighbors, which causes visible holes in the fluid surface in the front. Some regions in the back of the scene are also influenced and affected by this issue. Increasing the number of iterations to $N = 6$ ensures a much smoother fluid surface, as well as the absence of holes. When using DTC and DTVS, the front of the scene looks smooth and evenly distributed compared to the reference simulation. In the case of DTC, smaller holes and a non-smooth fluid surface in the back are clearly visible. However, DTVS creates an overall appealing visualization, which also represents fine-grained details like the vortex in the front right.

Figure 9 (top left) shows the influence on the average density in percent across all particles in the simulation during the time of the selected images. All APBF graphs appear to be scaled and/or shifted versions of the reference simulation-density graph. The most significant deviation is around 4% compared to the reference algorithm.

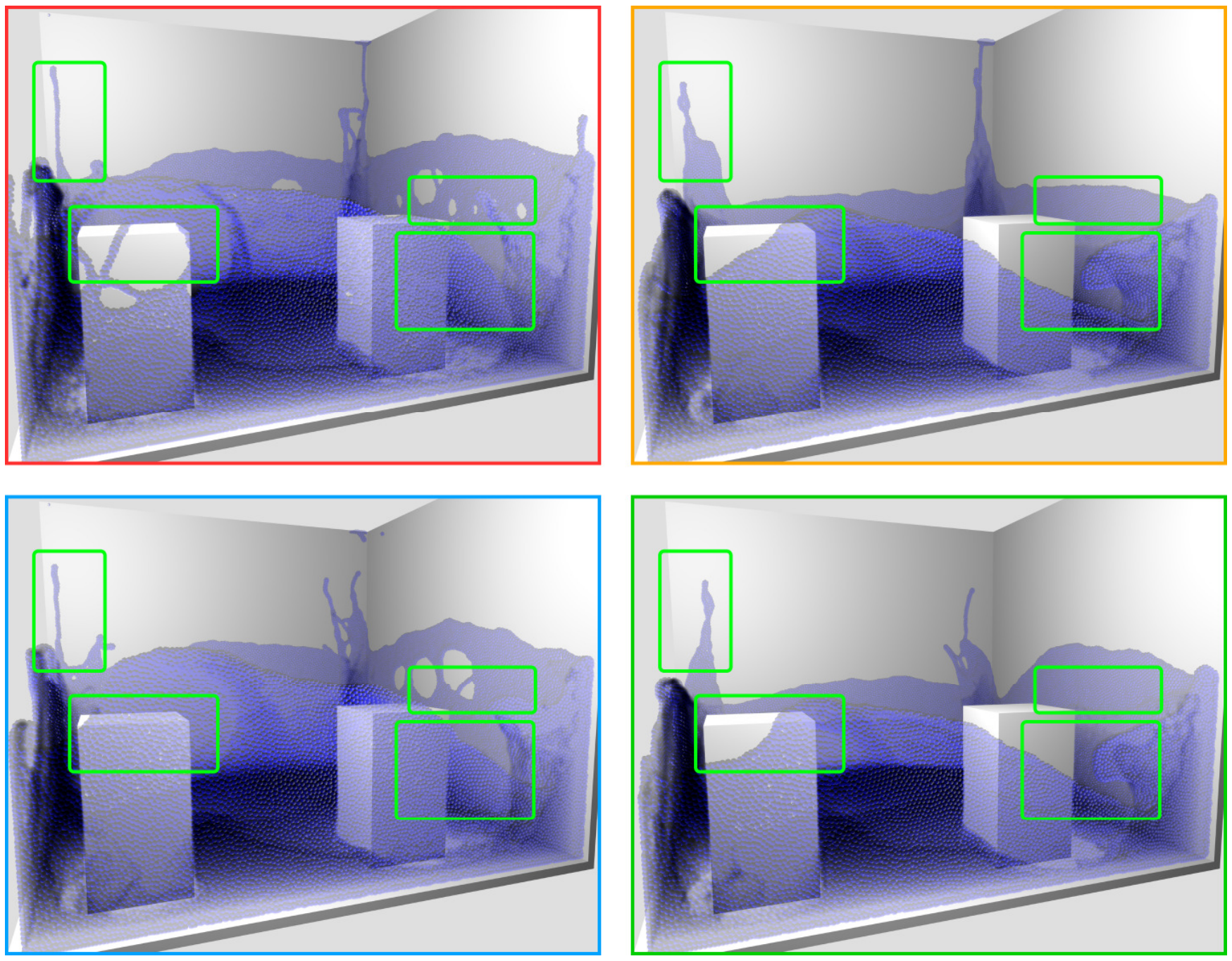

Figure 8. Comparison of the evaluation images for the dam-break scenario. Upper row: PBF with low quality (left, red, $N = 3$) and high quality (right, yellow, $N = 6$); lower row: APBF ($N \in \{3, \ldots, 6\}$) with DTC (left, blue) and DTVS (right, green). Differences between the methods are highlighted.

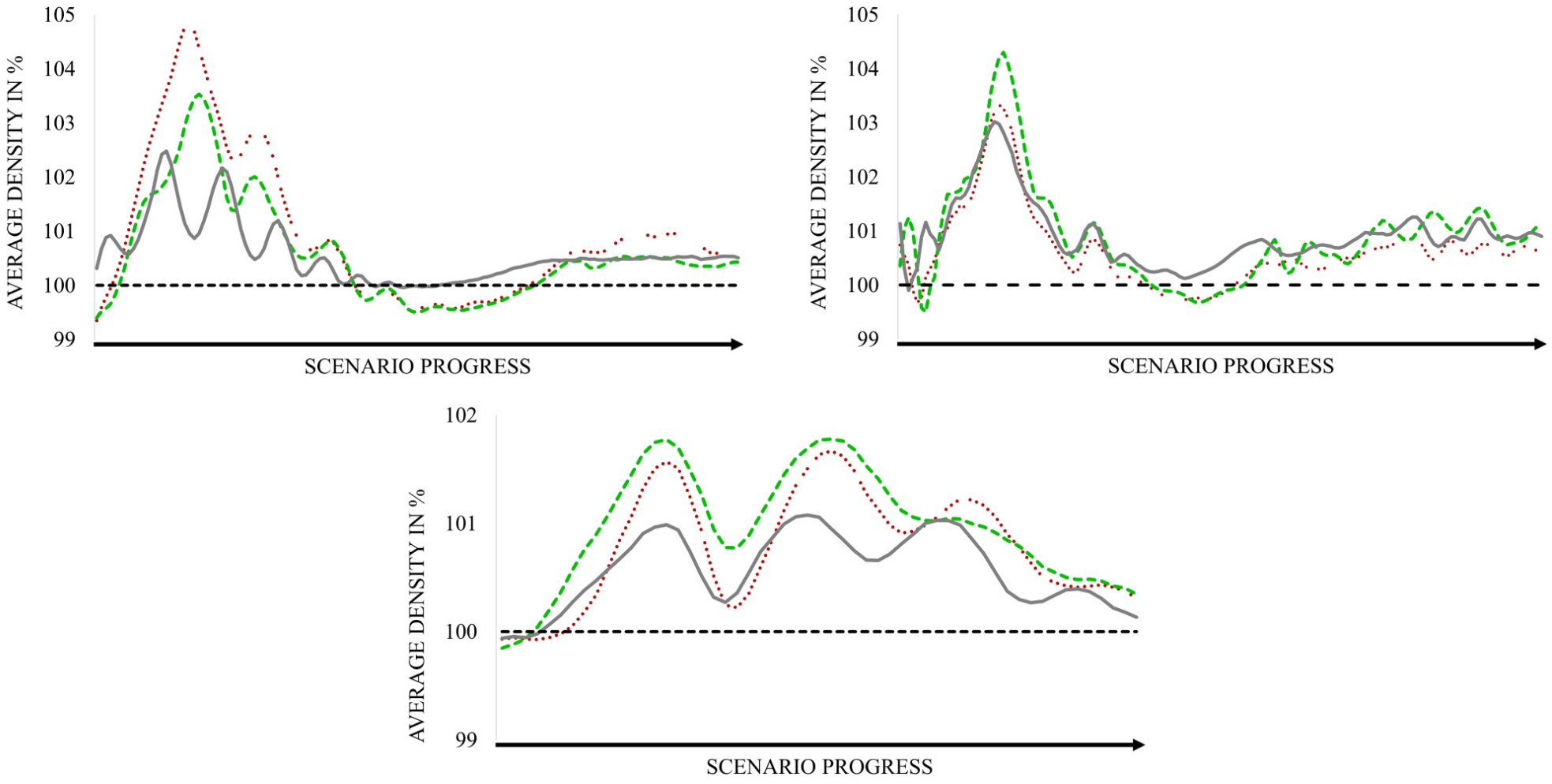

Figure 9. Most significant average-density deviations for the evaluation scenarios (limited amount of time). Our method has density deviations that are less than 4% on the evaluation scenarios. In general, APBF achieves comparable densities to PBF.Color coding: Black, rest density. Red, APBF with the DTC model (dotted). Green, APBF with the DTVS model (dashed). Gray, the original PBF approach (solid).

## 7.2 Double Dam Break

This setting uses two dam-break volumes of a high-density, inviscid fluid. It contains two dams with 336,400 particles each (= 672,800 particles) and uses $N = 5$ and $N = 10$ for the PBF versions.

Concerning visual quality, the two PBF fluid configurations in Figures 1 and 10 can be easily distinguished from each other in both cases: The low LOD version of the PBF method (with $N = 5$) shows unevenly distributed particles at the boundaries.Furthermore, the central intersection volume contains hole-like structures due to an insufficient approximation of the densities.The higher-quality version (with $N = 10$) ensures a smooth-looking simulation and does not suffer from such artifacts. The adaptive versions with DTC and DTVS realize a high-quality simulation in the front. Again, the DTVS configuration achieves the best results in terms of visual quality compared to the reference version with $N = 10$.

The measured density deviations are below 1.5% in this case (Figure 9 top right).This deviation was measured during the intersection of both fluid volumes and represents the worst case in this scenario. As before, the graphs of our method appear to be similar compared to the reference method.

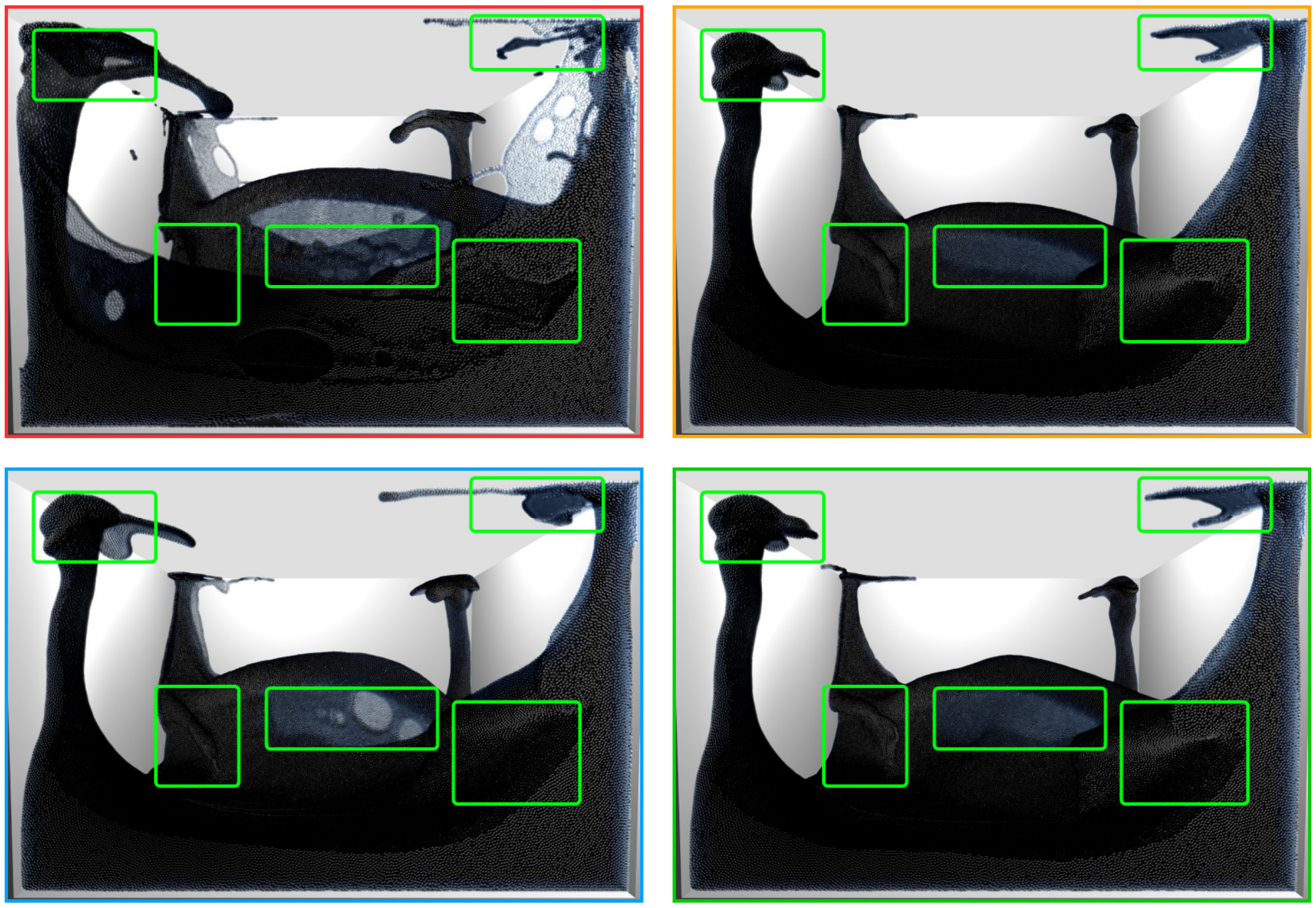

Figure 10. Comparison of the evaluation images for the double-dam-break scenario. Upper row: PBF with low quality (left, red, $N = 5$) and high quality (right, yellow, $N = 10$); lower row: APBF ($N \in \{5, \ldots, 10\}$) with DTC (left, blue) and DTVS (right, green). Differences between the methods are highlighted.

### 7.3. Multi Dam Break

Four small initial volumes with 56,350 particles each that sum up to a number of 225,400 particles in total are used in this case (Figure 11).

The fluid is configured as a medium-density and highly viscous fluid that uses $N = 4$ and $N = 8$ for the PBF versions. The chosen rendered frame for this scenario presents the collision between the walls, the central cone and the fluid volumes. Increasing the number of PBF iterations from $N = 4$ to $N = 8$ results in a better simulation of the modeled fluid.In particular, the fluid peaks in the front are preserved. All adaptive approaches reach an overall good visual quality: The results are very similar to that of the reference version with $N = 8$.

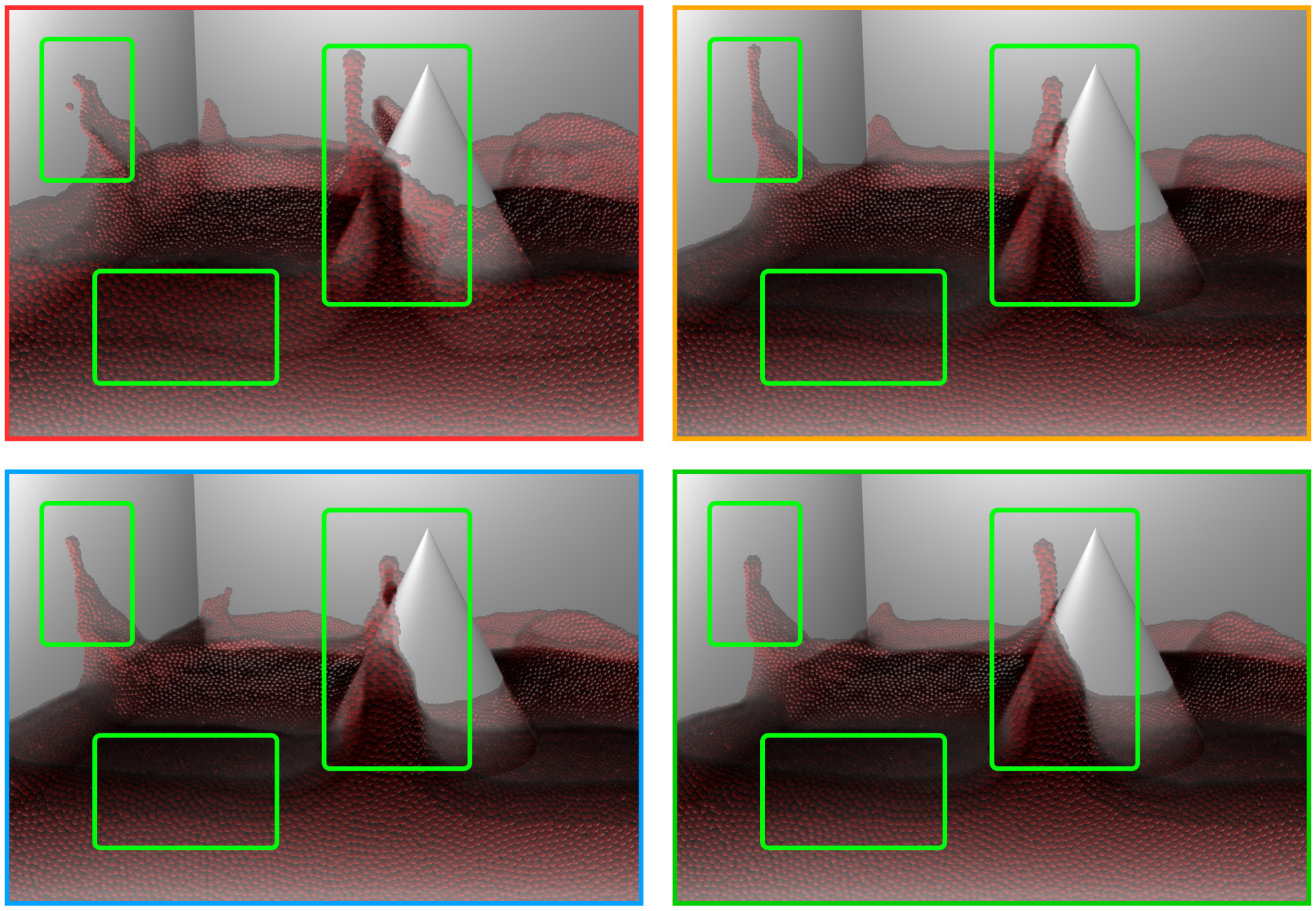

Figure 11. Comparison of the evaluation images for the multi-dam-break scenario. Upper row: PBF with low quality (left, red, $N = 4$) and high quality (right, yellow, $N = 8$); lower row: APBF ($N \in \{4, \dots, 8\}$) with DTC (left, blue) and DTVS (right, green). Differences between the methods are highlighted.

Figure 9 (bottom) displays the average-density measurement during the time of the analyzed frame. Similar to the previous scenario, the density deviations are around 1.5% in the worst case.

## 7.4. Performance

As presented in Table 1, the performance degrades noticeably when increasing the number of iterations per frame on both GPUs.Comparing the original PBF method against our approach reveals that we are able to increase the performance on every scenario by at least 40%.The measured maximum improvement in performance was 77% on the GeForce GTX 680 and 69% on the Radeon HD7850.Overall, the DTVS adaption model performs slightly better compared to the DTC model in terms of performance in image quality.

However, the benefit depends on the scenario and the adaption-model configurations used. In our cases, we used settings that preserve high-quality simulation results. When trading quality against performance, the runtime speed can be further improved.

## 8. Conclusion and Future Work

We presented a novel approach for adaptive fluid simulations using position-based constraints. It is a lightweight extension to the method by [4] that can be easily included in an existing position-based solver. Using LOD-based particle sets, we are able to achieve significant improvements in performance of up to 77% on the evaluation scenarios, in comparison to the reference method. In terms of visual quality, the results remain nearly the same using the adaptive concept. Comparing the two evaluated adaption models, namely DTVS and DTC, DTVS results in a better simulation quality and runtime performance. Our method is also capable of maintaining a similar average density of the simulated fluids compared to PBD.Other more sophisticated adaption models for the computation of LOD information might also be beneficial. However, changing the adaption model to assign LOD in a non-linear way might break the preservation of the average-density and stability.

In the current algorithm, we only consider additional contact constraints that are also solved adaptively in a particle-focused way. In the future, we would like to investigate other adaptive approaches, such as adaptive particle sizes. Furthermore, integration into a unified solver, e.g. a parallel successive over-relaxation solver [7], might reveal new insights.

## Acknowledgements

We would like to thank Nane Neu for her support and advises related to images and design. Furthermore, we would like to thank Michael Schmitz.